\documentclass[pre, twocolumn,
eqsecnum,
amsmath,amssymb,
]{revtex4}

\newcommand{\be}{\begin{equation}}
\newcommand{\ee}{\end{equation}} 
\newcommand{\bea}{\begin{eqnarray}} 
\usepackage{graphicx}
\newcommand{\eea}{\end{eqnarray}}
\usepackage{epsfig}
\usepackage{bm}
\usepackage[dvips]{color}

\begin{document}

\title{Option Pricing from Wavelet-Filtered Financial Series}
\author{V.T.X. de Almeida and L. Moriconi\footnote{Corresponding author. tel: +55 21 25627917. 
\\ E-mail address: moriconi@if.ufrj.br}}
\affiliation{Instituto de F\'\i sica, Universidade Federal do Rio de Janeiro, \\
C.P. 68528, 21945-970, Rio de Janeiro, RJ, Brazil}
\begin{abstract}
We perform wavelet decomposition of high frequency financial time series into large 
and small time scale components. Taking the FTSE100 index as a case study, and working 
with the Haar basis, it turns out that the small scale component defined by most 
($\simeq$ $99.6 \%$) of the wavelet coefficients can be neglected for the purpose of 
option premium evaluation. The relevance of the hugely compressed information provided 
by low-pass wavelet-filtering is related to the fact that the non-gaussian statistical structure
of the original financial time series is essentially preserved for expiration times which are larger
than just one trading day.
\\

\end{abstract}
\maketitle

\section{Introduction}

The problem of option pricing  \cite{bouch-pott,mant-stan,voit} has been a main 
topic of investigation in much of the econophysics literature, challenged by the 
well-known inadequacy of the standard Black-Scholes model to the real world 
\cite{bouch-pott,mant-stan,voit, matacz,borland,lmori}. Options are an imperative 
element in modern markets, since they play a fundamental role, as convincingly 
shown long ago by Black and Scholes, in reducing portfolio risk. As an alternative to 
the Black-Scholes model, one of the authors has implemented an option pricing scheme 
which is based on the evaluation of statistical averages taken over samples generated 
from the underlying asset log-return time series \cite{lmori}. This method, which we 
will refer to as ``Empirical Option Pricing" (EOP), has been succesfully validated through 
a careful study of FTSE100 options. 

A deeper understanding of the statistical features of financial time series is in order, since 
this would eventually allow us to replace real samples by accurate synthetic financial series, 
improving the statistical ensembles used in EOP. As a closely related issue, our aim in this work 
is to show that financial series can be hugely compressed (we mean lossy compression, in the 
information theoretical sense) by wavelet-filtering, without spoiling option premium evaluation 
by EOP. The low-pass wavelet-filtered signal contains log-return fluctuations defined on time 
scales larger than a few hours and it is likely to yield, due to its high compression rate, a 
more suitable ground for modeling and synthetization.

This paper is organized as follows. Secs. II and III provide brief accounts, respectively, of 
the EOP method and of the low-pass wavelet-filtering procedure that has been applied to our 
analysis of the FTSE100 index. The wavelet-filtered financial series, which keeps only $0.04 \%$ 
of the total number of wavelet components of the original signal is noted to encode the essential 
statistical information needed for a consistent evaluation of FTSE100 option premiums with 
expiration times larger than a single trading day. In Sec. IV, we summarize our findings and point 
out directions of further research.
 
\section{Empirical Option Pricing (EOP)}

We rephrase here, without paying much attention to rigorous considerations, the main points 
of EOP \cite{lmori}. Let $S(t)$ be an arbitrary financial index modeled as a continuous stochastic 
process. More precisely, we write down a Langevin evolution equation for $S(t)$, which is a simple 
generalization of the one underlying the Black-Scholes model \cite{hull, wilm_etal}:
\be
\frac{d S}{dt} = \mu(t) S + \sigma(t) \eta(t) S \ . \ \label{s_eq}
\ee
Above, $\mu(t)$ and $\sigma(t)$ are the time-dependent interest rate and volatility 
of the index $S$. The stochasticity of the financial time series comes from the 
gaussian white noise process $\eta(t)$ appearing in Eq. (\ref{s_eq}), which satisfies to
\bea
&&\langle \eta(t) \rangle = 0 \ , \ \nonumber \\
&&\langle \eta(t) \eta(t') \rangle = \delta(t-t') \ . \
\eea
Observe that both $\mu(t)$ and $\sigma(t)$ may be regarded as stochastic process
as well, with fluctuations correlated on time scales which are much larger than the 
correlation time of $S(t)$.

Working within the It\^o prescription, Eq. (\ref{s_eq}) can be readily rewritten
as
\be
\frac{dx}{dt} = \sigma(t) \eta(t) \ , \ \label{dxdt}
\ee
where
\be
S(t) = S \exp \left [ \int_0^t dt'\tilde \mu(t') + x(t)-x(0) \right ] \ , \ \label{st}
\ee
with
\be
\tilde \mu(t) \equiv \mu(t) - \frac{1}{2} \sigma(t)^2 \ . \  \label{eq_mu-mu}
\ee
Above, $S \equiv S(0)$ is just the spot price of the index. We are interested, now, 
to evaluate the premium of an european option which is negotiated with strike price $E$ 
and expiration time $T$. Similarly to what is done in the Black-Scholes treatment, 
where $\mu(t)$ and $\sigma(t)$ are constant, the option premium $V$ (for, say, call options) 
can be obtained from the computation of the statistical average
\be
V = \exp[-rT] \langle (S(T)-E) \Theta(S(T)-E) \rangle \ , \ \label{eq_v}
\ee
where $\mu(t)$ is replaced by $r$, the risk-free interest rate, in 
the definition of $S(T)$ provided by Eqs. (\ref{st}) and (\ref{eq_mu-mu}).

For stochastic processes $\{x_n\}$ defined in discrete time, with time step $\epsilon$, 
like real financial time series, Eq. (\ref{dxdt}) can be replaced by the finite difference 
equation
\be
\frac{1}{\epsilon} (x_{n+1} - x_n) = \sigma_n \eta_n \ , \  \label{dxdt2}
\ee
where 
\be
\eta_n \equiv \frac{\xi_n}{\sqrt{\epsilon}}
\ee
and $\xi_n = \pm 1$ is an arbitrary element of a discrete gaussian stochastic process, 
defined by $\langle \eta_n \rangle =0$ and $\langle \eta_n \eta_m \rangle = \delta_{nm}$. 
From (\ref{dxdt2}), we get, immediately,
\be
\sigma_n^2 = \frac{1}{\epsilon} (x_{n+1} - x_n)^2 \equiv \frac{1}{\epsilon}(\delta x_n)^2
\ee
and, therefore,
\be
\frac{1}{2} \int_0^T dt \sigma(t)^2 \simeq \frac{1}{2} \sum_{n=0}^{N-1} 
(\delta x_n)^2 \ , \ 
\ee
where time instants are given by $t_n = n\epsilon$, with $T= N \epsilon$.

Substituting, now, (\ref{st}) and (\ref{eq_mu-mu}) (with $\mu(t)$ replaced by $r$)
in (\ref{eq_v}),
we get
\bea
S(T)&=& S \exp \left [ r T +x(T)-x(0) - \frac{1}{2} \sum_{n=0}^{N-1} 
(\delta x_n)^2 \right ]  \nonumber \\
&=& S \exp \left [ r T +\sum_{n=0}^{N-1} \left ( \delta x_n - \frac{1}{2} (\delta x_n)^2 \right ) \right ]  \ . \ \label{st2}
\eea
It is important to note that $\delta x_n$, which appears in the above expressions 
is, from Eq. (\ref{st}), nothing more than an element of the detrended log-return series,
i.e.,
\be
\delta x_n = \ln [S(t_{n+1})/S(t_n)] - \epsilon \tilde \mu(t_n) \label{dx} \ , \
\ee
where $\langle \delta x \rangle =0$, due to Eq. (\ref{dxdt}). 

We are now ready to summarize EOP in the four following steps:

(i) A large period ($>$ two years) of reasonably statistically stationary high-frequency (minute-by-minute) 
log-return series of the underlying asset is ``purified" by the remotion of outlier events (typically, 
log-return fluctuations which are larger than 10 standard deviations) and of the mean one-week asset's 
interest rate (detrending). The resulting series is a stochastic process $\{ \delta y_n \}$;

(ii) Since the historical volatility $\sigma = \sqrt{ \langle (\delta y_n)^2 \rangle }$ is in general different from 
the volatility of the financial series during the option lifetime $T$, we introduce a correction factor $g$ to 
define the stochastic process $\{\delta x_n = g \delta y_n\}$, which yields a putative volatility
$\sigma^* = g \sigma$ for that period \cite{comment}.  The $g$-factor is the only adjustable parameter in EOP, which 
accounts for the distinction between the past and the future behaviors of the financial index $S(t)$.

(iii) An ensemble $\cal{E}$ of samples, each of length $T=N \epsilon$ ($\epsilon = 1$ minute) is defined from 
one-hour translations of the initial sequence $\{\delta x_0$, $\delta x_1$, ..., $\delta x_{N-1} \}$. In other words,
\be
{\cal{E}} = \bigcup_{m, \Delta} \{ \delta x_{m \Delta}, 
\delta x_{1 + m \Delta},...,\delta x_{N -1+ m \Delta}  \}  \ , \  \label{E}
\ee
where $m \in \mathbb{N}$  and $\Delta = 60$;

(iv) Option premiums are computed from (\ref{eq_v}), (\ref{st2}) and (\ref{dx}), 
with statistical averages taken over the ensemble $\cal{E}$, defined in (\ref{E}).
We note, furthermore, that the optimal value for the $g$-factor is found through the least squares method, devised 
for the comparison between the market and modeled option premiums.

A good agreement has been attained between the market and EOP values in a detailed study of the FTSE100 index 
options \cite{lmori}. The comparison data is reported in the MKT and OP columns of Table I.

The performance of EOP would benefit greatly from the use of synthetic financial series which would 
enlarge the ensemble of samples $\cal{E}$. Thus, one may wonder, having modeling aims in mind, on what are 
the relevant statistical facts hidden in the financial time series. The essential question we address is, accordingly, 
whether the financial series be decomposed into relevant and irrelevant contributions, as far as option pricing 
is concerned. In the next section, we recall some ideas on wavelet-filtering, which have been crucial in the 
investigation of this issue.

\section{Wavelet-Filtering and EOP}

\begin{figure}[tb]\includegraphics[width=11.0cm, height=15.0cm]{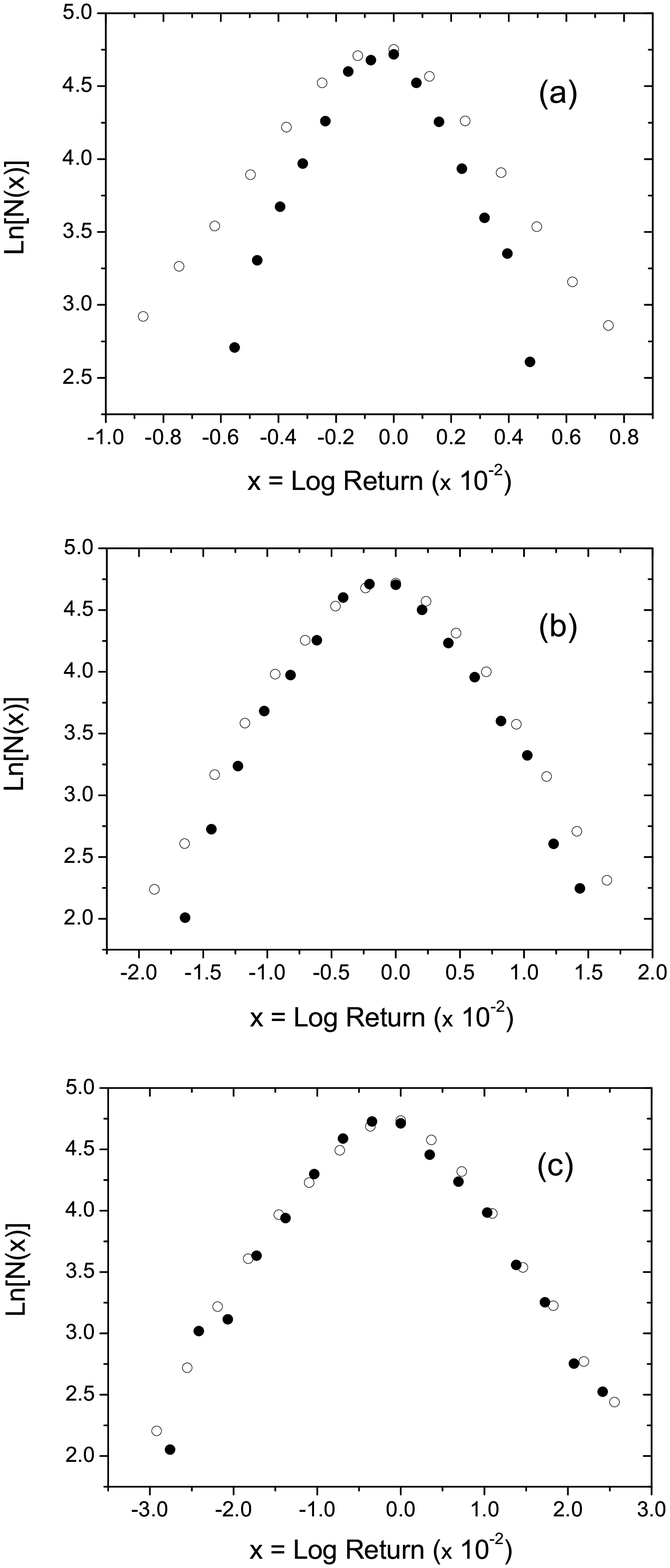}
\caption{Monolog plots of the histograms of log-return fluctuations for both the original signal 
(empty circles) and the wavelet-filtered series (filled circles) taken for time horizons of (a) 100 minutes,
(b) 300 minutes, and (c) 600 minutes. We call attention to the non-gaussian profiles of theses 
distributions.}
\label{}
\end{figure}

Log-return fluctuations are constantly affected by avalanches of market orders which have to do with
speculative trends, and are clearly time-localized events. These features render the financial time series 
suggestively adequate for wavelet analysis.

Since there is no requirement of continuity for the log-return time series, we have chosen, due to 
easiness of handling, to work with Haar wavelets \cite{walker}. In the same way as for any other 
discrete wavelet basis, the Haar wavelets are labelled by two integer indices $1 \leq j \leq J$ and 
$0 \leq k \leq 2^j-1$ and are given by
\be
\psi_{jk}(t) = \psi_{00}(2^jt-k) \ , \
\ee
where
\be
\psi_{00}(t) = \left\{ \begin{array}{rcl}
1 & \mbox{for} & 0 \leq t < \frac{1}{2}  \\
-1 & \mbox{for} &  \frac{1}{2} \leq t < 1   
\end{array}\right.
\ee
is the function known as ``mother wavelet". Observe that the above basis functions 
are defined in the domain $0 \leq t < 1$.

\begin{center}
\begin{table*}[tb]
\begin{tabular}{|c|c|c|c||c|c|c||c|c|c|}
\hline
{Strike}
&\multicolumn{3}{c||}{02dec05 ($S=5528.1$) }&\multicolumn{3}{c||}{06dec05 ($S=5538.8$)}&\multicolumn{3}{c|}
{09dec05 ($S=5517.4$)}\\
\cline{2-10}
{Price}
&MKT&OP$$& ${\overline{\hbox{OP}}}$& MKT&OP&${\overline{\hbox{OP}}}$&MKT&OP& ${\overline{\hbox{OP}}}$ \\\hline
$5125$  & 410.5   & 412.67  &  413.00  & X     & X      & X           & X     & X     & X          \\
$5225$  & 312     & 312.79  & 313.12   & 324   & 321.87 & 322.25  & 298   & 297.51 & 297.74  \\
$5325$  & 214.5   & 213.94  & 214.19   & 225.5 & 222.87 & 223.15  & 199   & 197.72 & 197.87  \\
$5425$  & 122.5   & 121.93  & 122.17   & 131.5 & 129.48 & 129.65  & 103.5 & 102.35 & 102.20   \\
$5525$  & 50      & 48.61   & 48.55    & 53.5  & 53.52  & 53.41    & 29.5  & 29.72  & 29.15      \\
$5625$  & 13      & 13.01   & 13.00      & 12.5  & 14.97  & 14.90     & 3.5   & 4.79   &  4.65   \\
$5725$  & 2.5     & [0.60]  & 0.60       & 2     & 1.66   & 1.65       & 0.5   & 0.35   & 0.29       \\
$5825$  & 0.5     & [0.0]   & 0.0       & X     & X      & X          & X     & X       &    X      \\
\hline
\hline
{Strike}
&\multicolumn{3}{c||}{19dec05 ($S=5539.8$)}&\multicolumn{3}{c||}{03jan06 ($S=5681.5$)}&\multicolumn{3}{c|}{12jan06 ($S=5735.1$)} \\
\cline{2-10}
Price&MKT&OP&${\overline{\hbox{OP}}}$& MKT&OP&${\overline{\hbox{OP}}}$&MKT& OP&${\overline{\hbox{OP}}}$ \\\hline
$5225$  & 329.5   & 332.38  & 333.05  & X     & X      & X      & X    & X       & X       \\
$5325$  & 234.5   & 237.49  & 237.64  & 368.5 & 368.36 & 368.84 & 414   & 414.96 & 415.12  \\
$5425$  & 148     & 149.46  & 150.09  & 271   & 268.86 & 269.30 & 314   & 315.02 & 315.18  \\
$5525$  & 76      & 77.75   & 77.97   & 177   & 176.67 & 174.97 & 215   & 215.09 & 215.25  \\
$5625$  & 28.5    & 30.91   & 31.09   & 93    & 91.20  & 91.40  & 119   & 116.81 & 116.81  \\
$5725$  & 8       & [5.18]  & 5.09    & 34.5  & 34.40  & 34.44  &  40   & 34.72  & 34.38   \\
$5825$  & 2.5     & [0.60]  & 0.54    & 9     & 8.41   & 8.39   &  5.5  & 4.51   & 4.36    \\
$5925$  & 0.5     & [0.0]   & 0.0     & 2     & [0.19] & 0.18   & 0.5   & 0.20   & 0.16    \\ 
$6025$  & X       & X       &  X      & 0.5   & [0.0]  & 0.0    & X     &  X     & X       \\
\hline
\end{tabular}

\caption{Call option premiums taken from the market (MKT) are listed together with the EOP evaluations performed with the original (OP) and
the wavelet-filtered (${\overline{\hbox{OP}}}$) series. The market data was recorded on 02dec05, 06dec05, 09dec05, 19dec05, 
03dec06, and 12dec06; spot prices are indicated by $S$; the respective g-factors (and putative volatilities; see Sec. II) are $g=0.81$ ($\sigma^*= 6.1 \%$), $g = 0.91$ ($\sigma^* = 6.9 \%$), $g = 0.94$ ($\sigma^* = 7.1 \%$), $g=0.78$ ($\sigma^*= 5.9 \%$), $g = 0.83$ ($\sigma^* = 6.3 \%$), and 
$g = 0.78$ ($\sigma^* = 5.9 \%$). The risk-free interest rate is $r=4.5\%$. The three first dates in december refer to options which expired
on 16dec05. For the other three dates, expiration date is 20jan06. The mean volatilities measured between 02dec05 and 09dec05 and between 
19dec05 and 12jan06 were $\sigma = 8.0 \%$ and $\sigma = 6.1 \%$, respectively. The brackets in some of the OP$_0$ evaluations indicate 
option premiums which are probably affected by poor sampling of the underlying financial series.}
\end{table*}
\end{center}

The detrended log-return series $\{ \delta x_0,  \delta x_1, ...,  \delta x_{N-1} \}$ of 
length $N=2^{J+1}$ and zero mean \cite{comment2} can always be expanded in wavelet 
modes as 
\be
\delta x_i = \sum_{j=0}^J \sum_{k=0}^{2^j-1} c_{jk} \psi_{jk}(i/N) \ . \  \label{wav_exp}
\ee
Low-pass wavelet-filtering can be straightforwardly implemented from the expansion 
(\ref{wav_exp}) by retaining the modes which have the scale index $j < j^\star$, where
$j^\star$ is an arbitrarily fixed threshold.  We have taken (following the prescriptions 
given in step (i) of EOP, as discussed in Sec. II) a financial time series of 241664 minutes 
(around two years of data) for the the FTSE100 index, ending on 17th november, 2005. 
The series is partitioned into 59 subseries, each of length 4096 (corresponding to $J=11$ 
and about two weeks of market activity), which are then wavelet-filtered with threshold
 parameter $j^\star = 4$ (compression rate of $99.6 \%$). 

Since wavelets with $j>0$ have zero mean, we expect that the histograms of log-returns 
$ \sum_{n=0}^{N-1} \delta x_i$ will not be much affected for time horizons 
$T = N \epsilon  > 4096/2^{j^\star} = 256$ minutes. This is actually verified in Fig. 1. Therefore,  
on the grounds of Eq.  (\ref{st2}), it is clear that option prices can be alternatively estimated 
through the use of the low-pass wavelet filtered series within  EOP for time horizons which are 
larger than one trading day ($T=510$ minutes).

In Table I, we report the computed premiums for call options based on the FTSE100 index with expiration 
times ranging from a few days to one month, in december 2005 and january 2006. The agreement between
the original and the wavelet-filtered option premium evaluations is significative. It is important to recall,
as already indicated in Ref. \cite{lmori}, that the Black-Scholes framework is unable to yield
good estimates of the market option premiums listed in Table I.

\section{Conclusions}

We have found, taking the FTSE100 index as a case study, that its high frequency 
(minute-by-minute) time series can be highly compressed for the purpose of option pricing. 
The original and the low-pass wavelet-filtered series have remarkably similar performances 
within EOP, even for a compression rate of $99.6\%$, which means that only 967 out of the
original 241664 wavelet components have been selected throught the wavelet-filtering 
procedure. The retained wavelet coefficients have scale index $j$  smaller than the 
fixed threshold $j^\star=4$ and are associated to log-return fluctuations defined on time
scales larger than a few hours. It turns out, thus, that one is entitled to use the filtered time 
series to precify FTSE100 options with expiration times which are larger than just one trading day, 
where log-returns are still clearly non-gaussian random variables. A  promising approach to 
option pricing, deserved for further investigation, is to address the problem of series synthetization 
from the analysis of the statistical properties of the compressed wavelet-filtered financial indices 
directly in wavelet space, in a spirit similar to what is done in the context of artificial multrifractal 
series \cite{arneodo_etal2}. It is likely that EOP would, then, be considerably improved from 
the use of much larger synthetic statistical ensembles.

\acknowledgements
This work has been partially supported by CNPq and FAPERJ.
The authors would also like to thank an anonimous referee for
a number of insightful comments, which have been of fundamental
importance in shaping the final form of this work.


\begin{references}

\bibitem{bouch-pott} J.P. Bouchaud and M. Potters, {\it{Theory of Financial Risks 
- From Statistical Physics to Risk Management}}, Cambridge University Press, Cambridge (2000).

\bibitem{mant-stan} R. Mantegna and H.E. Stanley, {\it{An Introduction to Econophysics}}, 
Cambridge University Press, Cambridge (2000).

\bibitem{voit} J. Voit, {\it{The Statistical Mechanics of Financial Markets}}, 
Springer-Verlag (2003).

\bibitem{matacz} A. Matacz, Int. J. Theor. Appl. Finance {\bf{3}}, 143 (2000).

\bibitem{borland} L. Borland, Phys. Rev. Lett. {\bf{89}}, 098701 (2002); 
Quant. Fin. {\bf{2}}, 415 (2002).

\bibitem{lmori} L. Moriconi, Physica A {\bf{380}}, 343 (2007).

\bibitem{hull} J. Hull, {\it{Options, Futures and Other Security Derivatives}},
Prentice Hall, New Jersey (1993).

\bibitem{wilm_etal} P. Wilmott, S. Howison, and J. Dewinne, 
{\it{The Mathematics of Financial Derivatives}}, Cambridge University Press, 
Cambridge (1995).

\bibitem{comment} Whenever volatilities are given in percentual amounts, 
throughout the paper, we mean the ``annualized volatilities", defined
by $\sigma_0 \sqrt{252 \times 60 \times 8.5}$, where $\sigma_0$ is the 
standard deviation of the minute-by-minute log-return financial series 
(we assume 252 trading days per year, and 8.5 market hours per day).

\bibitem{walker} J.S. Walker, {\it{A Primer on Wavelets and Their
Scientific Applications}}, 
Chapman $\&$ Hall/CRC (1999).

\bibitem{comment2} Our real detrended samples have some residual
mean values, which are, in practice, neglegible for all purposes
in EOP (see the second column in table III). In the general 
case, an additional wavelet basis function $\phi(t) =1$, 
for $0 \leq t \leq 1$, is included in the expansion 
(\ref{wav_exp}).

\bibitem{arneodo_etal2} A. Arn\'eodo, E. Bacry, and J.-F. Muzy,
J. Math. Phys. {\bf{39}}, 4163 (1998).

\end{references}
\end{document}